\documentstyle[aps,multicol,psfig,epsf,epsfig]{revtex}            

\begin{document}
\def\pa{\parallel}
\def\pe{\bot}
\let\a=\alpha \let\b=\beta  \let\c=\chi \let\d=\delta  \let\e=\varepsilon
\let\f=\varphi \let\g=\gamma \let\h=\eta \let\k=\kappa  \let\l=\lambda
\let\m=\mu   \let\n=\nu   \let\o=\omega    \let\p=\pi
\let\r=\varrho  \let\s=\sigma \let\t=\tau   \let\th=\vartheta
\let\y=\upsilon \let\x=\xi \let\z=\zeta
\let\D=\Delta \let\F=\Phi  \let\G=\Gamma  \let\L=\Lambda \let\Th=\Theta
\let\O=\Omega       
\draft
\tightenlines
\title{Dynamic behavior of  anisotropic  non-equilibrium 
driving lattice gases.}
\author{Ezequiel V. Albano and Gustavo Saracco}
\address{Instituto de Investigaciones Fisicoqu\'{\i}micas Te\'oricas y 
Aplicadas (INIFTA), UNLP, CONICET, Casilla de Correo, 
16 Sucursal 4, (1900) La Plata, Argentina}
\date{\today}

\maketitle
\begin{abstract}
It is shown that intrinsically anisotropic non-equilibrium systems
relaxing by a dynamic process exhibit universal critical behavior
during their evolution toward non-equilibrium stationary
states. An anisotropic scaling anzats for the dynamics is proposed 
and tested numerically.
Relevant critical exponents can be evaluated
self-consistently using both the short- and long-time dynamics 
frameworks. The obtained results allow us to 
clarify a long-standing
controversy about the theoretical description, the universality and the
origin of the anisotropy of driven diffusive systems, 
showing that the standard field theory does not hold and supporting 
a recently proposed alternative theory. 
\end{abstract}


\begin{multicols}{2}
\narrowtext
The development of a theoretical framework for the study
of non-equilibrium systems 
is of vital importance for a wide range of disciplines in both
science and technology. In order to overcome
such a difficult problem, it is useful to study models
capable of capturing the essential physical features of the
real system. The statistical mechanics of
interacting lattice gases, driven into
non-equilibrium steady states (NESS) by external fields \cite{zia,MD},
has attracted growing attention. 
Some remarkable characteristics of the NESS observed in 
driven diffusive systems (DDS) are, among others, 
their non-Hamiltonian nature, the violation of the fluctuation 
dissipation theorem and the fact that the steady-state 
distribution is determined by the dynamics. 

The prototypical DDS has early been proposed
by Katz, Lebowitz and Spohn (KLS) \cite{katz} as a
generalization of the Ising model.
Then, the KLS model is a kinetic Ising model with conserved dynamics.
Using the lattice gas language, particle hopping in the direction 
(against the direction) of an externally applied field ($\vec{E}$) 
is favored (unfavored) \cite{zia,MD,katz}, while jumps perpendicular
to the field are unaffected by it.
For half-filled lattices and high 
enough temperatures the KLS model exhibits a disordered phase.
However, at low temperatures an anisotropic ordered NESS 
emerges and is characterized by strips of high particle
density crossing the lattice in the direction parallel to 
the applied drive \cite{zia,MD}. 
So, at some critical temperature ($T_{c}$) the
KLS model undergoes a second-order non-equilibrium phase transition. 
The issue of the universality class of 
the KLS model is of general interest because it will contribute
to the rationalization of non-equilibrium critical phenomena. 
However, this task  
is elusive and has become the subject of a long-standing debate.
A detailed discussion of such conceptually rich  
debate is beyond the aim of this work.
Instead, we will briefly comment their theoretical and numerical aspects.

{\it The theoretical controversy }. Aimed to analyze the critical 
nature of DDS and determine their
universality, a Langevin equation
%, which is assumed to be 
%a coarse-grained version of the discrete model, 
has been proposed  \cite{JS}. 
This equation describes the stochastic evolution of the local 
particle density $\rho({\bf x},t)$ and in terms of $\phi = 2\rho -1$,
it reads
\begin{eqnarray}
\label{dds}
\partial_t\phi({\bf r})= &&
-\nabla_{\pe}^4\phi+\tau_\pe \nabla_{\pe}^2\phi
+{g\over6}\D_{\pe}\phi^3
 \nonumber \\
&& + \tau_\pa  \nabla_{\pa}^2 \phi
- \epsilon  \nabla_{\pa} \phi^2
+ {\bf \z}({\bf r},t),
\end{eqnarray}  

\noindent where ${\bf \z}({\bf r},t)$ is a conserved noise term that 
reflects the fast degrees of freedom, $\epsilon$ 
denotes the coarse-grained drive, 
and $\tau_{\pe}$, $\tau_{\pa}$ and $g$  are model parameters.
This field theory leads to anisotropic scaling 
wave vectors in the critical region, with
non-trivial correlation length exponents $\nu_{\pe}$ and $\nu_{\pa}$. 
The current term  ($\epsilon  \nabla_{\pa} \phi^2$)
is the most relevant nonlinearity
and source of anisotropy.

Another Langevin equation, which is a coarse-grained 
version of the master equation, has also been proposed \cite{gallegos}.
Although such equation has been criticized \cite{crit},
after healing some flaws\cite{healing}, the new 
equation, which is valid 
in the limit of an infinite driving field, reads
\begin{eqnarray}
\label{ads}
\partial_t\phi({\bf r})= &&
-\nabla_{\pe}^4\phi+\tau_\pe \nabla_{\pe}^2\phi
+{g\over6}\D_{\pe}\phi^3
 \nonumber \\
&& + \tau_\pa  \nabla_{\pa}^2 \phi
+ {\bf \z}({\bf r},t) .
\end{eqnarray}     

Comparing Eqs.(1) and (2) it follows that they only differ 
in the current term $\epsilon  \nabla_{\pa} \phi^2$
that vanishes in (Eq.(2)). 
This difference has deep consequences, including: 
i) The critical dimension of Eq.(2) is $d = 3$ instead of  $d = 5$  
in Eq.(1). ii) Renormalization group analyses give 
different critical exponents (see Table I).
iii) It has been argued that the most 
relevant non-equilibrium effect is the anisotropy
generated by the driving field \cite{gallegos1} 
and not the current \cite{zia}.
iv) In Eq.(2), the current term vanishes in the infinite 
driving limit, as shown rigorously in \cite{gallegos}. 

Another piece of heavy debate is the {\it disagreement in the 
interpretation and evaluation of numerical data}. 
Valles and Marro \cite{vama}, have concluded 
that $\beta$ seemed to be close to $\beta \approx 0.3$,
as predicted by Eq.(2). In view of the 
anisotropic nature of the system, Leung \cite{ktl} has developed an 
anisotropic finite-size approach. Extensive computer
simulation \cite{ktl,wan} results give data collapsing when the shape of 
the lattice ($S = L_{y}^{\nu_{\pe}/\nu_{\pa}}/L_{x} $) 
and $\beta$ are taken in accordance with 
Eq.(1). However, a subsequent analysis of Leung's data \cite{ktl}, 
discussed in \cite{MD}, shows that they may also be consistent 
with $\beta$ close to $0.3$. Also recent simulations
give $\beta \approx 0.33$ \cite{gallegos1}.

We note that the analysis of numerical data obtained under 
NESS conditions may be biased 
by some assumptions such as  anisotropic {\it vs} isotropic scaling, 
the shape $S$ of the lattice,
the method used to estimate $T_{c}$ in the thermodynamic limit,
the quality of data collapsing, etc.
   
%Within this context, 
The aim of this work is to provide an unambiguous clarification 
of the existing controversy, based 
on extensive numerical studies of both the short- and 
the long-time dynamics of four variants of the KLS model. 
In their work, Janssen et al. \cite{JSstd} have shown 
that a system relaxing by a dynamic process exhibits universal behavior
at the early stages of its evolution towards {\bf equilibrium} states. 
This idea has been verified in many 
models exhibiting critical behavior under 
{\bf equilibrium} conditions \cite{BZ}.
Since the extension of this concept to the 
field of {\bf non-equilibrium} critical behavior and to 
{\bf anisotropic systems} has not been performed yet, 
the present work has additional ingredients of general interest.

The KLS model \cite{katz} is defined on the  
square lattice assuming a rectangular
geometry $L_x, L_y$, using periodic boundary conditions.
A lattice configuration is specified by the set of
occupation numbers $n_{i,j} = \{0,1\}$, corresponding to each
site of coordinates $(i,j)$. 
Nearest-neighbor (NN) attraction ($J > 0$) 
is considered. So, in the absence of 
a field the Hamiltonian $\cal H$ is given by
\begin{equation}
{\cal H} = -4J \sum_{<ij,i'j'>} n_{i,j} n_{i',j'}       , \label{ham}
\end{equation}
\noindent where the summation runs over NN sites only.
The driving field $E$ acts along the $y-$direction. 
The coupling to a thermal bath at temperature $T$
and the action of the field are
considered through Metropolis rates, i.e.
$min [ 1, exp -( \{\Delta {\cal H} - \eta E\} /k_{B}T )$,
where $k_{B}$ is the Boltzmann constant, 
$\Delta \cal H$ is the change in $\cal H$
after the exchange, and $ \eta = (-1,0,1)$ for
a particle attempting to hop (against, orthogonally, along)
the field, respectively.

Monte Carlo simulations are performed using lattices of 
different sizes, with  $240 \leq L_{x} \leq 960$ and 
$30 \leq L_{y} \leq 480$. $T$ is reported in units of 
$J/k_{B}$ and $E$ is given in units of $J$. 
In all cases the density
%of particles 
is $\rho_{o} = 1/2$.
%  remains constant.
One Monte Carlo time step (mcs) involves
$L_{x} \times L_{y}$ trials. 

Four variants of the KLS model are studied, namely:
{\it i}) the infinite driving limit (IKLS) with $E = \infty$,
{\it ii}) a finite driving case (FKLS) with $E = 0.5$,
{\it iii}) the random field KLS model \cite{RKLS} in the
infinite driving limit (IRKLS) and 
{\it iv}) the oscillatory KLS model \cite{OKLS} in the 
infinite driving limit (IOKLS). In the IRKLS (IOKLS), 
the driven field takes values $E = \pm \infty$ at 
random (with a period $\tau_{o}$), 
generating anisotropy but not an overall
current. So, the study of the IKLS model 
is aimed to determine the validity of 
either Eq.(1) or Eq.(2). Also, the FKLS is studied to
check the validity of Eq.(1) in the finite driving case. 
Finally, the comparison of  the results of the 
IKLS model and those of both the  IRKLS and the IOKLS models
will help to clarify  the role of the current. 
 
In order to observe universality in the dynamics the system has 
to be initialized with configurations far from the NESS.
For this purpose two different initial states have been 
employed, namely:
i) Fully disordered configurations (FDC) as expected 
for $T \rightarrow \infty$ and
ii) The ground state configuration (GSC) as expected at $T = 0$,
which is a single strip parallel to the drive \cite{muka}.

The order parameter ($OP$) is defined 
as the excess density in the direction parallel to the
applied field, namely
\begin{equation}
OP \equiv (R L_{x})^{-1}
\sum_{i=1}^{L_{x}} |P(i) -\rho_{o}|   ,  \label{op1}
\end{equation}
\noindent where $P(i) = (L_{y})^{-1} \sum_{j=1}^{L_{y}} n_{ij}$
is the density profile along
the $x-$direction (perpendicular to the drive) 
and $R = (2\rho_{o}(1 - \rho_{o}))$ is
a normalization constant. For simulations starting from the
GSC the OP given by Eq.(4) and that proposed by Leung \cite{ktl}
give the same results. However, in contrast to Leung's OP,
that given by Eq. (4) is suitable to detect the onset of 
multi-stripped ordering during the short-time dynamics when 
simulations start from the FDC. Since $OP$ involves the 
calculation of the absolute value of the
excess density, it is not suitable for the observation of  
the expected critical initial increase in the order 
parameter \cite{BZ}. Instead, 
$OP$ can be identified with the square root of the second moment
of the excess density. So,   
using FDC starting configurations, the proposed scaling anzats, 
which generalizes the standard one \cite{BZ}, reads:
\begin{eqnarray}
OP(t,\tau,L_{y},L_{x}) = b^{-\beta/\nu_{\pa}} OP^{*}(b^{-z}t, 
b^{1/\nu_{\pa}}\tau, \nonumber \\ 
b^{-1}L_{y}, b^{-\nu_{\pe}/\nu_{\pa}}L_{x}),
\end{eqnarray}
\noindent where $\tau = T - T_{c}/T_{c}$, $OP^{*}$ is a 
scaling function, $b$ is the spatial rescaling factor,
$\beta$ is the order parameter critical
exponent, $\nu_{\pa}$ ($\nu_{\pe}$) is the correlation length 
exponent in the direction parallel (perpendicular) to the drive, 
and $z$ is the dynamic critical exponent, respectively.
In order to generate the FDC the lattice is filled at random
with probability $p = 1/2$. While a procedure is used to ensure that 
the whole density of the sample is exactly $\rho_{o} = 1/2$, 
density fluctuations of the order of $L_{y}^{-1/2}$ 
are present along the columns parallel to the drive. 
These tiny fluctuations add up according to Eq.(4), so that
the amplitude of the $OP$ depends on $L_y$ as $L_{y}^{-1/2}$. 
Taking  $b \propto t^{1/z}$ in Eq.(5), it follows 
\begin{equation}
OP(t,L_{y}) \propto L_{y}^{-\frac{1}{2}} t^{c_{2}} \nonumber \\  
OP^{**}(t^{1/\nu_{\pa} z}\tau), L_{x},L_{y} \rightarrow \infty,
\end{equation}
\noindent with  $c_{2} = (1 - 2 \beta/\nu_{\pa})/2z$ \cite{BZ}. 

Figure 1(a) shows log-log plots of $OP(t) L_{y}^{1/2}$  {\it vs} $t$
for the IKLS model obtained at $T = 3.20$. 
It should be noticed that a tiny 
downward (upward) deviation from linearity is obtained 
for  $T = 3.21$ ($T = 3.19$) (not shown here). 
So, our estimation of the critical 
temperature is $T_{c}^{IKLS} = 3.20 \pm 0.01$. So, 
$T_{c}^{IKLS} \approx 1.41 T_{c}^{I}$, where  $T_{c}^{I}$ is the Onsager 
temperature of the $d = 2$ Ising model. Notice that this value
is fully consistent with previous 
estimations \cite{gallegos1,vama,ktl,wan}.
Data collapsing of the raw data already observed for 
different lattice sizes supports the anzats of Eq.(6), and this
result confirms the  assumption, also implicit in Eq.(6), that 
for the case of FDC the only relevant length of the lattice 
is that parallel to the drive along which the precursors of 
the striped patterns start to develop. The best fit of the data 
shown in Figure 1(a) gives $c_{2} = 0.114 \pm 0.005$. Notice that 
all the critical exponents evaluated in this work are listed 
in Table I. 
\begin{figure}
\centerline{
\epsfxsize=8cm
\epsfysize=6cm
\epsfbox{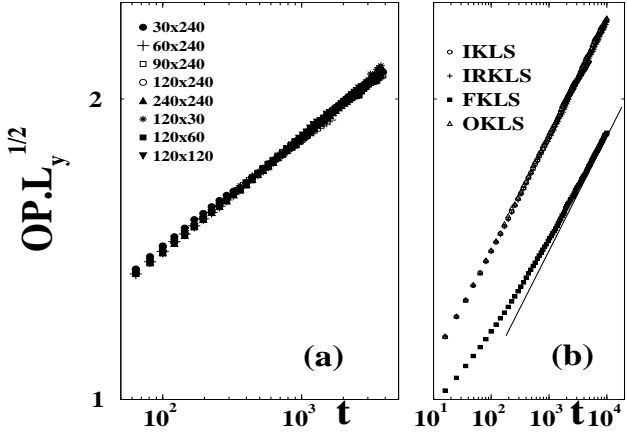}}
\caption{Log-log plots of the scaled order parameter {\it vs} $t$
at criticality. Results averaged over $10^4$ different realizations.
(a) Data corresponding to the IKLS model obtained starting 
with FDC and using lattices of different
sizes as shown in the figure. 
(b) Data corresponding to the four studied models and
obtained using lattices of size $L_{x} = 240, L_{y} = 120$.
The straight line has been drawn to show that after some
transient, the curve of the FKLS model becomes
almost parallel to those of the other cases.}
\label{Figure1}
\end{figure} 
Figure 1(b) allows the comparison of data obtained for the
four different models studied. A perfect 
overlap of the data corresponding to the IKLS, 
the IRKLS ($T_{c}^{IRKLS} = 3.16 \pm 0.02$) and the 
OKLS ($T_{c}^{OKLS} = 3.16 \pm 0.02$) is found,
leading to almost the same exponent $c_{2}$, as shown in Table I.
Data of the FKLS model ($T_{c}^{FKLS} = 2.78 \pm 0.015, E = 0.5$)
exhibit a crossover from a very early
behavior with appreciable curvature to a latter
regime ($t \geq 10^{3}$) where the slope of the 
other cases is recovered (within error bars, see Table I).     

Furthermore, taking the logarithmic derivative 
of $OP$ given by Eq. (6) at criticality, one has \cite{BZ}
\begin{equation}
\partial{ln} OP(t,\tau) |_{\tau = 0} \propto t^{c_{\pa}}  , 
c_{\pa} = \frac{1}{\nu_{\pa}z}.
\end{equation}
\noindent The best fits of the data give the values of the
exponents $c_{\pa}$ listed in Table I.

Starting from GSC's, we propose that the standard scaling behavior 
of $OP$ 
can be generalized as\cite{BZ}:
\begin{eqnarray}
OP(t,\tau,L_{x},L_{y}) = b^{-\beta/\nu_{\pe}} 
OP^{**}(b^{-z}t, b^{1/\nu_{\pe}}\tau, \nonumber \\ 
b^{-\nu_{\pa}/\nu_{\pe}}L_{y}, b^{-1}L_{x}),
\end{eqnarray}
\noindent where $OP^{**}$ is a scaling function. 
Taking $b \approx t^{1/z}$  in Eq.(8) at criticality ($\tau = 0$), 
it follows that:
\begin{equation}
OP(t) \propto  t^{-\beta/\nu_{\pe} z}.
\end{equation}
Also taking the derivative of Eq.(8) with respect to $\tau$ 
it follows that
\begin{equation}
\partial OP(t,\tau) |_{\tau = 0} \propto t^{c_{\pe}}  , 
c_{\pe} = \frac{1}{\nu_{\pe} z} - \frac{\beta}{\nu_{\pe}z}.
\end{equation}
Figure 2(a) shows log-log plots of $OP(t)$ {\it vs} $t$,
for the IKLS model obtained for $T = 3.200$.
Notice that starting from a GSC, the power law
decay of the OP as described by Eq.(9) is obtained
after a long time, e.g. for $t \geq 10^{5}$ in Figure 2(a). 
A detailed view of this behavior is shown in Figure 2(b)
for the four cases studied. Also, fitting these curves
using Eq.(9) the slopes given by $\beta/\nu_{\pe} z $ can 
be obtained, as listed in Table I.

It should be noticed that  tiny 
downward and upward deviations from linearity are observed 
for temperatures slightly out of criticality (not shown here).
However, using GSC, this method for the location 
of $T_{c}$ is roughly twice less sensitive 
than that used starting with FDC.  
So, our second (independent)  estimation of the critical 
temperature is $T_{c}^{IKLS} = 3.20 \pm 0.02$.
This value is fully consistent with 
our previous estimation made starting from FDC,
as expected for second-order transitions. The same behavior is 
observed in the other cases studied.
Data collapsing of the raw data as observed in Figure 2(a)
supports the assumption, implicit in the proposed scaling anzats  
(Eqs. (8-9)), that starting from 
a GSC the only relevant length of the lattice is that 
perpendicular to the drive along which a 
diffusion-like process causes relaxation to the NESS.
Our results confirm that the dynamics does not reveal the anisotropy 
of the system through the shape of the 
lattice $S = L_{y}^{\nu_{\pe}/ \nu_{\pa}}/L_{x}$,
in contrast to the scaling treatment of 
data obtained under NESS conditions \cite{ktl,wan,binwan}.
Instead, the anisotropy enters through the different correlation 
length exponents used in the scaling relationships, i.e.,
Eqs.(6) and (9). 

Log-log plots of the derivative of Eq.(8) (not shown here) 
support the scaling anzats of  Eq.(10) and allow
the evaluation of $c_{\pe}$, as listed in Table I. 

Combining the exponents $\beta/\nu_{\pe}z$ and $c_{\pe}$
obtained starting with GSC it is possible to evaluate
$\beta$. Also, using those exponents, as well as $c_{2}$
and $c_{\pa}$ as evaluated starting from FDC, it is possible 
to calculate $\nu_{\pe}$, $\nu_{\pa}$ and $z$ (see Table I).
Furthermore, for anisotropic systems such as the KLS model, the
following modified hyperscaling relation is expected 
to hold \cite{binwan}:
\begin{equation}
\nu_{\pa} + (d - 1)\nu_{\pe} -2\beta = \gamma,
\end{equation}
\noindent where $\gamma$ is the exponent of the susceptibility.
So, using the numerically evaluated exponents, our estimations of
$\gamma$ made with the aid of Eq.(11)  are also listed in Table I.
\begin{figure}
\centerline{
\epsfxsize=8cm
\epsfysize=6cm
\epsfbox{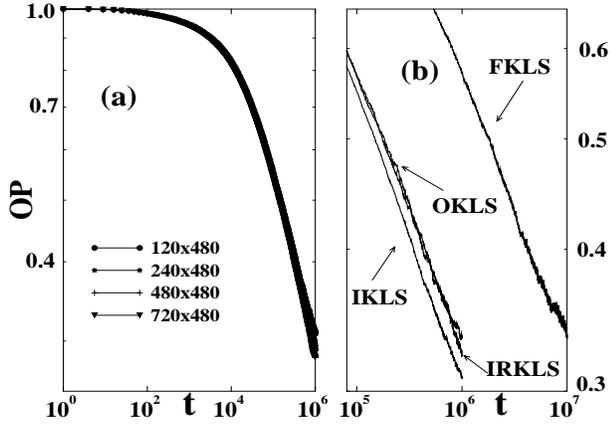}}
\caption{Log-log plots of the order parameter {\it vs} $t$ as 
obtained starting with GSC.
a) Results corresponding to the IKLS model using lattices of different
sizes as shown in the figure. 
b) Results obtained for the different models studied as identified
in the figure. Data obtained using lattices of size
$L_{x} = 480$, $L_{y} = 240$, except for the FKLS model with
$L_{x} = 960$, $L_{y} = 60$.}
\label{Figure2}
\end{figure} 
Based on the results summarized in Table I, 
it is concluded that: 
i) All critical exponents can be obtained
self-consistently within the dynamics framework. 
In contrast to previous numerical estimations performed
under NESS conditions, the values of the exponents are {\bf not-biased } 
by any assumption on the shape of the lattice used. 
ii) All the evaluated exponents are consistent with Eq.(2),
and the predictions of Eq.(1) are far outside our error bars. 
iii) Since two models with macroscopic current (IKLS, FKLS) and 
another two without it (IRKLS, IOKLS), have the same critical
exponents, it follows that such current is not relevant neither
to determine the universality class nor for the emergence of 
the anisotropy.   
iv) The short-time dynamics of the studied non-equilibrium
systems exhibits universal critical behavior, as already
observed in equilibrium systems.

%In summary, this work helps to clarify the ongoing
%debate on the theoretical framework, the universality and
%the origin of the anisotropy, in driving diffusive systems.

{\bf Acknowledgments}: This work was supported by CONICET, 
UNLP and ANPCyT (Argentina). G.S. acknowledges the CIC for 
a research fellowship.

\vspace{-0.25cm}
\begin{table}
\caption {List of critical temperatures and  exponents.
The symbols (*), (+)  and (*+) stand for exponents 
obtained using data corresponding to FDC and GSC,
and combinations of both, respectively.
The predictions of both Eq. (1) and Eq.(2), up to the first 
term of the $\epsilon$ expansion, are also 
listed. The results of the oscillatory models are less sensitive
to tiny changes of $T$, causing the exponents 
to be less accurate than in the other cases.
Notice that $z$ is renormalizated with respect to the 
perpendicular direction.} 
\end{table}
\end{multicols}
\vspace{0.3cm}
{\centering \begin{tabular}{|c|c|c|c|c|c|c|c|c|c|c|}
\hline 
\(MODEL \)&
\(T_{c} \)&
\( c_{2} (*)\)&
\( c_{\pa} (*)\)&
\( \beta/\nu_{\pe} z (+) \)&
\( c_{\pe} (+) \)&
\( \beta  (+)  \)& 
\( \nu_{\pe} (*+) \)&
\( \nu_{\pa} (*+) \)&
\( z (*+) \)&
\( \gamma \) \\
\hline 
\hline 
IKLS&
3.200(10)&
0.114(5)&
0.406(10)&
0.254(15)&
0.516(15)&
0.330(30)&
0.644(40)&
1.221(40)&
2.016(40)&
1.21(11)   \\
\hline 
\hline 
FKLS&
2.780(15)&
0.108(5)&
0.409(10)&
0.245(15)&
0.487(15)&
0.335(30)& 
0.668(40)&
1.198(40)&
2.041(40)&
1.20(11)  \\
\hline 
\hline 
IRKLS&
3.160(20)&
0.115(5)&
0.421(10)&
0.228(15)&
0.506(15)&
0.311(30)&
0.671(40)&
1.168(40)&
2.033(40)&
1.22(11) \\
\hline 
\hline 
IOKLS&
3.160(20)&
0.116(5)&
0.424(10)&
0.235(15)&
0.505(15)&
0.281(30)&
0.635(40)&
1.110(40)&
2.126(40)&
1.18(11) \\
\hline 
\hline 
Eq. (1)&
---&
$\approx 1/8$ &
$\approx 1/2$ &
$\approx 3/4$ &
$\approx 3/4$ &
$1/2$ &
$\approx  1/2$&
$\approx 3/2$&
$\approx 4/3$&
$\approx 1$ \\
\hline 
\hline 
Eq. (2)&
---&
$ \approx 0.114$&
$ \approx 0.410$&
$ \approx 0.262$&
$ \approx 0.532$&
$ \approx 0.33$ &
$ \approx 0.63$ &
$ \approx 1.22$ &
$ \approx 1.998$&
$ \approx 1.17$\\
\hline 
\end{tabular}\par}
\vspace{0.3cm}

%\begin{multicols}{1}

%\end{multicols}


\begin{thebibliography}{99}
\vspace*{-1.75cm}

\bibitem{zia} B. Schmittmann and R. K. P. Zia, {\em Statistical Mechanics of
Driven Diffusive Systems\/}, in Phase Transitions and Critical
Phenomena, edited by C. Domb and J. Lebowitz (Academic, London, 1995)

\bibitem{MD} J. Marro and R. Dickman,  {\it Nonequilibrium Phase Transitions
in Lattice Models
}, Cambridge University Press, (Cambridge, U.K., 1999).

\bibitem{katz} S. Katz, J.L. Lebowitz and H. Spohn,
Phys. Rev. B {\bf 28}, 1655 (1983); J. Stat. Phys. {\bf 34}, 497 (1984).

\bibitem{JS} H. K. Janssen and B. Schmittmann, Z. Phys. B {\bf 64}. 503
(1986). K.-t. Leung and J. L. Cardy, J. Stat. Phys. {\bf 44}, 567
(1986); ibid {\bf 45}, 1087 (Erratum) (1986).

\bibitem{gallegos} P. L. Garrido, F. de los Santos and M. A. Mu\~noz,
Phys. Rev. E {\bf 57}, 752 (1998). 

\bibitem{crit} B. Schmittmann, et al., Phys. Rev. E, {\bf 61}, 5977 (2000).

\bibitem{healing} P. L. Garrido, M. A. Mu\~noz, and F. de los Santos,
Phys. Rev. E {\bf 61}, R4683 (2000). 
 
\bibitem{gallegos1} A. Achahbar, P. L. Garrido, J. Marro and M. A. Mu\~noz,
Phys. Rev. Lett. {\bf 87}, 195702 (2001).

\bibitem{vama} J. L. Vall\'es and J. Marro, J. Stat. Phys. {\bf 49}, 89 (1987).

\bibitem{ktl} K.-t. Leung, Phys. Rev. Lett. {\bf 66}, 453 (1991).

\bibitem{wan} J. S. Wang, J. Stat. Phys. {\bf 82}, 1409 (1996).

\bibitem{JSstd} H. K. Janssen, B. Schaub and B. Schmitmman,
Z. Phys. {\bf 73}, 539 (1989).

\bibitem{BZ} B. Zheng. Int. J. Mod. Phys. B. {\bf 14}, 1419 (1998),
(Review article). H. J. Luo, L. Schuelke and B. Zheng, 
Phys. Rev. Lett. {\bf 81}, 180 (1998). B. Zheng, M. Schulz and S. Trimper, 
Phys. Rev. Lett. {\bf 82}, 1981 (1999).

\bibitem{RKLS} B. Schmittmann and R. Zia, Phys. Rev. Lett. {\bf 66},
357 (1991). 

\bibitem{OKLS} R. Monetti and E. Albano. Europhys. Lett. {\bf 56}, 
400 (2001). 

\bibitem{muka} E. Levine, Y. Kafri and D. Mukamel. Phys. Rev. E.
{\bf 64}, 026105 (2001).

\bibitem{binwan} K. Binder and J. S. Wang. J. Stat. Phys. {\bf 55}, 87 (1989).

\end{thebibliography}
\end{document}